\documentstyle[12pt]{article}
\begin{document}
\def\hw{\hbar \omega}
\def\lsim{\:\raisebox{-0.5ex}{$\stackrel{\textstyle<}{\sim}$}\:}
\def\gsim{\:\raisebox{-0.5ex}{$\stackrel{\textstyle>}{\sim}$}\:}
\def\d{\dagger}
\def\nsp{\negthinspace}
\def\b#1#2{\left(\nsp \begin{array}{r}#1\\#2\end{array} \nsp \right)}
\def\be{\begin{equation}}
\def\ee{\end{equation}}
\def\br{\begin{eqnarray}}
\def\er{\end{eqnarray}}
\def\x{\times}
\def\go{\rightarrow  }
\def\dodt{{$\bullet $ }}
\def\bra#1{\langle #1|}
\def\ket#1{|#1 \rangle}
\def\bra#1{\langle #1|}
\def\ov#1#2{\langle #1 | #2  \rangle }
\def\L {{\Lambda}}
\def\T {{\Theta}}
\def\U {{\Upsilon}}
\def\G {{\Gamma}}
\def\O {{{\cal O}}}
\def\M {{{\cal M}}}
\def\N {{{\cal N}}}
\def\Q {{{\cal Q}}}
\def\S {{{\cal S}}}
\def\up{u_{p}}
\def\vp{v_{p}}
\def\un{u_{n}}
\def\vn{v_{n}}
\def\upp{u_{p'}}
\def\vpp{v_{p'}}
\def\unp{u_{n'}}
\def\vnp{v_{n'}}
\begin{titlepage}
\pagestyle{empty}
\baselineskip=21pt
\rightline{to be published in Phys. Lett. {\bf B}}
\vskip .2in
\begin{center}
{\large{\bf Projected Linear Response Theory for Charge-Exchange
Excitations and Double Beta Decay }} \end{center}
\vskip .1in
\begin{center}

F. Krmpoti\'{c}$^{\d}$ and A. Mariano$^{\d}$

{\small\it Departamento de F\'\i sica, Facultad de Ciencias Exactas,}\\
{\small\it Universidad Nacional de La Plata, C. C. 67, 1900 La Plata,
Argentina}

T.T.S. Kuo

{\small\it Physics Department, State University of New York at Stony Brook,}\\
{\small\it Stony Brook, NY 11794-3800, USA}

{\small\it and}

K. Nakayama

{\small\it Department of Physics and Astronomy,
University of Georgia, Athens, Georgia 30602, USA}\\
{\small\it and}\\
{\small\it Institut f\"{u}r Kernphysik,
Forschungszentrum J\"{u}lich, D-52428 J\"{u}lich, Germany}

\end{center}
\vskip 0.5in
\centerline{ {\bf Abstract} }
\baselineskip=18pt
The projected random phase approximation (PRPA) for charge-exchange
excitations is derived from the time-dependent variational principle.
Explicit results for the unperturbed energies (including the self-energy
corrections), the PRPA matrices, and the transition matrix elements are
presented. The effect of the projection procedure on the two-neutrino
$\beta\beta$ decay in $^{76}Ge$ is briefly discussed.

\vspace{0.5in}
\noindent
$^{\dagger}$Fellow of the CONICET from Argentina.
\end{titlepage}
\baselineskip=18pt

In a recent work \cite{krm} simple formulae have been designed for the
$0^+\rightarrow 0^+$ double beta decay ($\beta \beta$) matrix elements
within the quasiparticle random phase approximation (QRPA).
In particular, it has been shown that the exact calculations for the $2\nu$
amplitude, when evaluated with a zero-range force, can be nicely fitted by
a $(1,1)$-Pad\'{e} approximant of the form
\be
{\cal M}_{2\nu}\cong {\cal M}_{2\nu}(t=0)
\frac{1-t/t_0}{1-t/t_1},
\label{1}
\ee
where $t=v^{pp}_t/v^{pair}_s$ is the ratio between the spin-triplet strength
in the particle-particle channel and the spin-singlet strength in the pairing
channel, and $t_0$ and $t_1$ denote, respectively, the zero and the pole
of ${\cal M}_{2\nu}$ and are treated as free parameters.
In the same work it has been suggested that this result is of general
validity and that any alteration of the nuclear hamiltonian or the
configuration space cannot lead to a different functional dependence.
\footnote[1] {When the renormalization coupling constant  $g^{pp}$ is used
\cite{tub} an analogous expression to the eq. (\ref{1}) is valid
(with $g^{pp}$'s for $t$'s).}
Thus, in order to modify the theoretical predictions for the
$\beta \beta $ moments in a qualitatively way,  the QRPA itself has
to be altered. The simplest amendment is to set up a
particle number projection. This has been done by Faessler and collaborators
in several recent papers \cite{tub}.
They have also discussed the outcomes of the PRPA on both simple and double
beta decays. We present here  a PRPA formalism that differs in several
aspects from the one reported by them.

To evade the disadvantages inherent in the non-conservation of
particle number, besides the usual proton quasiparticle transformation
\[
{\bf a}_{\sf p} =
\left(u_p {\bf c}_{\sf p}-v_p {\bf c}_{\bar{\sf p}}^{\d}\right)\quad;\quad
{\bf a}_{\sf p}^{\d}=
\left(u_p {\bf c}_{\sf p}^{\d}-v_p{\bf c}_{\bar{\sf p}}\right),
\label{2}
\]
it is useful to  introduce the following {\it canonical} transformation
\footnote[2] { Note that the operation symbolized by star ($\star$) does not
affect the complex variable $z_{\rm p}$.}
\begin{equation}
{\bf d}_{\sf p} =\sqrt{\sigma_p}
\left(u_p {\bf c}_{\sf p}-v_pz_{\rm p} {\bf c}_{\bar{\sf p}}^{\d}\right)\,;\,
{\bf d}_{\sf p}^{\star} =\sqrt{\sigma_p}
\left(u_p{\bf c}_{\sf p}^{\d}-v_pz_{\rm p} {\bf c}_{\bar{\sf p}}\right)\,;\,
\sigma_p^{-1}=u_p^2+z_{\rm p}^2v_p^2,
\label{3}
\end{equation}
and likewise for neutrons.
Here  ${\bf c}_{\sf p}^{\d}$ $({\bf c}_{\sf p})$, with ${\sf p} \equiv
p,m_{p}$,
$p \equiv {n_{p}\; \ell_{p}\; j_{p}}$ and $m_{p} \equiv m_{j_{p}}$, are the
single particle creation (annihilation) operators for protons,
and ${\bf c}_{\bar{\sf p}}=(-1)^{p+m_{p}}{\bf c}_{p,-m_{p}}$.
The  vacuums for the proton operators ${\bf d_p}$ is  the generating function
\cite{bay,die}
\[
\ket{z_{\rm p}}=  z^{\hat{Z}/2}_{\rm p}\ket{BCS;{\rm p}}=
\prod_{{\sf p}>0}\left(u_p + z_{\rm p} v_p {\bf c}_{\sf p}^{\d}
{\bf c}_{\bar{\sf p}}^{\d}\right) \ket{p},\label{4}
\]
where $\hat{Z}$ is the particle number operator for protons, and $\ket{p} $ and
$\ket{BCS;p}$ represent, respectively, the particle and the BCS
vacuum for protons.
The Bogoljubov transformation (\ref{3}) has been introduced in the past by
Ottaviani and Savoia \cite{ott} to perform a number projection
in spaces of two or more BCS quasiparticles. Later, their method has been
used by several authors (e.g., ref. \cite{all}).

First, we derive the following relations for the matrix elements of the model
hamiltonian H:
\begin{eqnarray}
\bra {BCS}H \hat{P}_0\ket{BCS}&=& {\cal T} \bra{z}H\ket{z},\label{5}\\
\bra{BCS}[{\bf A}(pn\bar{J})H\hat{P}_{\mu}{\bf A}^{\d}(p'n'J)]^0\ket{BCS}
&=& {\cal T} \bra{z}[{\bf D}_{\mu}(pn\bar{J})H
{\bf D}_{\mu}^{\star}(p'n'J)]^0 \ket{z},\label{6}
\end{eqnarray}
where $\ket{BCS}\equiv \ket{BCS,p}\ket{BCS,n}$ and
$\ket{z}\equiv\ket{z_{\rm p}}\ket{z_{\rm n}}$.
The projection operators are: $\hat{P}_0=\hat{P}_Z \hat{P}_N$,
$\hat{P}_{+1}=\hat{P}_{Z+1} \hat{P}_{N-1}$, and
$\hat{P}_{-1}=\hat{P}_{Z-1} \hat{P}_{N+1}$, with $\hat{P}_{\N}$ being the
operator for projecting states of good particle number ${\N}$ \cite{rin}.
The transformation
\[
{\cal T}=
(\frac{1}{2\pi i})^2\oint \frac{dz_{\rm p}}{z_{\rm p}^{Z+1}}
\oint \frac{dz_{\rm n}}{z_{\rm n}^{N+1}},
\label{7}
\]
is a contour integration including the origin, and
\[
{\bf A}^{\d}(pnJ)= [{\bf a}_{\sf p}^{\d}{\bf a}_{\sf n}^{\d}]^J;\,\,
{\bf D}_{\mu}^{\star}(pnJ)=
\sqrt{\sigma_p\sigma_n} z_{\rm p}^{(1-\mu)/2} z_{\rm n}^{(1+\mu)/2}
[{\bf d}_{\sf p}^{\star}{\bf d}_{\sf n}^{\star}]^J,
\label{8}
\]
are the proton-neutron excitation operators in the spaces of the operators
{\bf a} and {\bf d}, respectively.

For the norms, we get
\be
\bra {BCS} \hat{P}_0\ket{BCS} \equiv N_0 ={\cal T} {\sf N}_0,
\label{12}
\ee
and
\be
\hat{J}^{-1} \bra{BCS}[{\bf A}(pn;\bar{J})\hat{P}_{\mu}{\bf A}^{\d}(pnJ)]^0
\ket{BCS} \equiv N_{\mu}(pn) ={\cal T} {\sf N}_{\mu}(pn),
\label{13}
\ee
where
\[
{\sf N}_0=\bra{z}z\rangle;\;\;
{\sf N}_{\mu}(pn) = z_{\rm p}^{1-\mu}z_{\rm n}^{1+\mu}
\sigma_p\sigma_n \bra{z}z\rangle.
\label{14}
\]
The one-body charge-exchange operators
\[
{\cal O}_{+1}(J)=\sum_{\sf pn}\bra{{\sf p}}{\rm O}(J)\ket{{\sf n}}
{\bf c}_{\sf p}^{\d}{\bf c}_{\sf n}; \,\,\,\,\,
{\cal O}_{-1}(J)=\sum_{\sf np}\bra{{\sf n}}{\rm O}(J)\ket{{\sf p}}
{\bf c}_{\sf n}^{\d}{\bf c}_{\sf p},
\label{9}
\]
when expressed in terms of the {\bf d} operators, and after neglecting the
scattering terms that never contribute within a two quasiparticle space, read
\[
\O_{\mu}(J)\doteq \sum_{pn}
[\Lambda_{\mu}^{0}(pnJ) {\bf D}_{\mu}^{\star}(pnJ)
+ \Lambda_{-\mu}^{0}(pnJ) {\bf D}_{-\mu}(pn\bar{J})],
\label{10}
\]
with
\begin{eqnarray*}
&&\Lambda_{\mu}^{0}(pnJ)=-\hat{J}^{-1} \bra{p}|{\rm O}(J)|\ket{n}
 \left\{
\begin{array}{l}
\up \vn\;\; \mbox{for}\;\mu =1 \;,\\
\un \vp\;\; \mbox{for}\;\mu =-1 \;.\\
\end{array}\right.
\label{11}
\end{eqnarray*}
For their matrix elements we get
\be
\bra {BCS}[{\bf A}(pn\bar{J})\hat{P}_{\mu}\O_{\mu'}(J)]^0\hat{P}_0\ket{BCS}
\equiv O_{\mu}(pnJ)\delta_{\mu,\mu'}
={\cal T} {\sf O}_{\mu}(pnJ)\delta_{\mu,\mu'},
\label{15}
\ee
with
\[
{\sf O}_{\mu}(pnJ) =
\bra{z}[{\bf D}_{\mu}(pnJ){\O}_{\mu}(J)]^0\ket{z}
= {\sf N}_{\mu}(pn) \Lambda_{\mu}^{0}(pnJ).
\label{16}
\]
Relations (\ref{5}), (\ref{6}), (\ref{12}) and (\ref{13}) reveal
that, at the level of the Tamm-Dancoff approximation (TDA), there is a one
to one correspondence between the energy spectrum of H within the projected
basis $\hat{P}_{\mu}{\bf A}^{\d}(pnJ)\ket{BCS}$ (relative to the ground
state $\hat{P}_0\ket{BCS}$, with energy
$E_0=\bra{BCS}H\hat{P}_0\ket{BCS}/N_0$),
and that within the basis ${\bf D}_{\mu}^{\star}(pnJ)\ket{z}$ (relative to the
ground state $\ket{z}$, with energy ${\sf E}_0=\bra{z}H\ket{z}/{\sf N}_0$).
As it is seen from (\ref{15}) an analogous correspondence also stands for the
transition matrix elements $O_{\mu}(pnJ)$ and ${\sf O}_{\mu}(pnJ)$.
Thus within the TDA our method consists in the following:
{\it after evaluating the matrix elements of H and $\O_{\pm 1}(J)$
within the d-space ($<D>$), the corresponding projected matrix elements
($<P>$) are obtained from the transformation: $<{\rm P}>={\cal T}<{\rm D}>$}.

Next, the above method is extended to the RPA by following the linear
response procedure developed by  Lane and Martorell \cite{lan}.
We introduce the Slater determinant
\be
\ket{\phi (z)}
= \prod_{pnJ\mu}(1+ \chi_{\mu}(pnJ) {\bf D}_{\mu}^{\star}(pnJ))\ket{z},
\label{17}
\ee
and if the quantities $\chi$ are assumed to be small, then, to second order
\be
\langle \phi(z)\ket{\phi(z)}={\sf N}_0+
\sum_{pnJ\mu}\left[\left|\chi_{\mu}(pnJ)\right|^2 {\sf N}_{\mu}(pn)
+\left|\chi_{-\mu}(pnJ)\right|^2 {\sf N}_{-\mu}(pn)\right].
\label{18}
\ee
The response of the system to an external time-dependent field
\[
\hat{\O}_{\mu}=\O_{\mu}e^{-i\omega t} +\O_{\mu}^{\d}e^{i\omega t},
\label{19}
\]
is derived from the time dependent variational principle, which for the
optimum Slater determinant solution of the perturbed problem gives
\[
\bra{\delta \phi(z)}H-{\sf E}_0+\hat{\O}_{\mu}-
i\hbar\frac{\partial}{\partial t} \ket{\phi(z)}=0.
\label{22}
\]
If $ \chi_{\mu}(pnJ)$ are varied, keeping only the linear terms and assuming
$\bra{z} \hat{\O}_{\mu}\ket{z}=0$, one gets
\begin{eqnarray*}
&&{\sf A}_{\mu}\chi_{\mu}+{\sf B}\chi^{\ast}_{-\mu}
+{\sf O}_{\mu}e^{-i\omega t}-
i\hbar \dot{\chi}_{\mu}{\sf N}_{\mu}=0,
\\
&&{\sf B}^{\ast}_{\mu}\chi_{\mu}
+{\sf A}_{-\mu}^{\ast}{\chi}^{\ast}_{-\mu}
+{\sf O}_{-\mu}^{\ast}e^{-i\omega t}
-i\hbar \dot{\chi}_{-\mu}^{\ast}{\sf N}_{-\mu}^{\ast}=0,
\label{24}
\end{eqnarray*}
where the submatrices are given by
\begin{eqnarray*}
{\sf A}_{\mu}(pn,p'n'J) & = &
\hat{J}^{-1} \bra{z}[{\bf D}_{\mu}(pn\bar{J})(H-{\sf E}_0 {\bf 1})
{\bf D}_{\mu}^{\star}(p'n'J)]^0 \ket{z}
\\
{\sf B}(pn,p'n'J) & = &
\hat{J}^{-1}
\bra{z}[{\bf D}_{-\mu}(pn\bar{J}) {\bf D}_{\mu}(p'n'J)H]^0 \ket{z}.
\label{25}
\end{eqnarray*}
Attempting a solution of the form
\be
{\bf{\chi}}_{\mu}=  X_{\mu} e^{-i\omega t};\;\;
{\bf{\chi}}_{-\mu}=  Y^{\ast}_{\mu}e^{i\omega t},
\label{26}
\ee
and performing the transformation $(A_{\mu}, B, N_{\mu}, O_{\mu})
={\cal T}({\sf A}_{\mu}, {\sf B}, {\sf N}_{\mu}, {\sf O}_{\mu})$, we obtain
\begin{equation}
\left(\begin {array}{cc}
A_{\mu}-\hw N_{\mu}{\bf 1} &  B \\
B^{\ast} & A_{-\mu}^{\ast}
+\hw N_{-\mu}^{\ast} {\bf 1} \\
\end {array} \right)
\b{X_{\mu}} {Y_{\mu}}= -\b{O_{\mu}} {O_{-\mu}^{\ast}}.
\label{27}
\end{equation}
The corresponding RPA eigenvalue problem is:
\begin{equation}
\left(\begin {array}{cc}
A_{\mu} &  B \\
B^{\ast} & A_{-\mu}^{\ast}
\end {array} \right)
\b{X_{\mu}(\alpha)} {Y_{\mu}(\alpha)}=E_{\mu}(\alpha)
\left(\begin {array}{cc}
N_{\mu}  &  0 \\
0 & -N_{-\mu}^{\ast}
\end {array} \right)
\b{X_{\mu}(\alpha)} {Y_{\mu}(\alpha)},
\label{28}
\end{equation}
with the eigenvectors normalized such as to obey the orthogonal relation:
\begin{eqnarray*}
&&X_{\mu}^{\d}(\alpha) N_{\mu} X_{\mu}(\beta)
-Y_{\mu}^{\d}(\alpha) N_{-\mu} Y_{\mu}(\beta)=
 \left\{
\begin{array}{l}
\delta_{\alpha\beta}\;\; \mbox{for}\;E_{\mu}(\alpha)>0\;,\\
-\delta_{\alpha\beta}\;\; \mbox{for}\;E_{\mu}(\alpha)<0\;,\\
\end{array}\right.
\label{29}
\end{eqnarray*}
while the closure relations are:
\begin{eqnarray*}
&&\sum_{E_{\mu}(\alpha)>0} [ X_{\mu}(\alpha) X^{\d}_{\mu}(\alpha)
- Y_{-\mu}^{\ast}(\alpha)\tilde{Y}_{-\mu}(\alpha)]N_{\mu}={\bf 1},
\\
&&\sum_{E_{\mu}(\alpha)>0} [ X_{\mu}^{\ast}(\alpha) \tilde{Y}_{\mu}(\alpha)
- Y_{-\mu}(\alpha)X_{-\mu}^{\d}(\alpha)] N_{-\mu}^{\ast}=0.
\label{30}
\end{eqnarray*}

We interpret $E_{\mu}(\alpha J)$ as the PRPA approximation for the exact
excitation energies from the ground state in the nucleus ($Z$, $N$) to the
states $\ket{\alpha J;\mu}$ in the neighboring odd-odd nuclei:
$E_{\mu}(\alpha J) =\bra{\alpha J;\mu}H\ket{\alpha J;\mu}_{PRPA}
-\bra{0}H\ket{0}_{PRPA}$.
Note that in contrast with the usual charge-conserving RPA problem, the
eigenvalues do not occur in pairs $\pm E_{\mu}(\alpha J)$. Besides, both
positive and the negative solutions are physically meaningful. For $\mu= \pm1$
the positive solutions describe  excitations in the $(Z\pm1, N\mp1)$ nuclei,
while the negative energy solutions represent the de-excitations in the
$(Z\pm1, N\mp1)$ nuclei, and the positive energy excitations
in the $(Z\mp1, N\pm1)$ nuclei as well. Thus only one RPA equation has to be
 solved
for the evaluation of single $\beta^-$ and $\beta^+$ transitions.
This is a well known feature  of the charge-exchange modes \cite{lan,boh,krm1}.

We can express the solution of eq. (\ref{27}) in terms of the eigenvectors
of (\ref{28}),
\[
\b{X_{\mu}} {Y_{\mu}}=-\sum_{\alpha}sgn( E_{\mu}(\alpha))
\b{X_{\mu}(\alpha)} {Y_{\mu}(\alpha)}
\frac{1}{E_{\mu}(\alpha)-\hw}
(X_{\mu}^{\d}(\alpha)\, Y_{\mu}^{\d}(\alpha))
\b{O_{\mu}} {O_{-\mu}^{\ast}},
\label{31}
\]
and hence from (\ref{17}), (\ref{18}) and (\ref{26}) we derive the
response function \cite{lan},
\[
R(\hw)\equiv -\frac{1}{2}
\frac{{\cal T}\bra{\phi(z)}\hat{\O}_{\mu}\ket{\phi(z)}}
{{\cal T}\langle \phi(z)\ket{\phi(z)}}=
\sum_{\alpha J}sgn( E_{\mu}(\alpha J))
\frac{ |\Lambda_{\mu}(\alpha J)|^2}{E_{\mu}(\alpha J)-\hw},
\label{33}
\]
where
\[
\Lambda_{\mu}(\alpha J)=N_0^{-1/2}\sum_{pn}[
\Lambda_{\mu}^0(pnJ) N_{\mu}(pn) X_{\mu}^{\ast}(pn;\alpha J)
+\Lambda_{-\mu}^{0}(pnJ) N_{-\mu}(pn) Y_{\mu}^{\ast}(pn;\alpha J)],
\label{34}
\]
is the PRPA approximation to the matrix element of $\O_{\mu} (J)$ between
the exact counterparts of $\ket{\alpha J;\mu}$ and the ground state, i.e.,
$\Lambda_{\mu}(\alpha J)= \bra{\alpha J;\mu}| \O_{\mu} (J)| \ket{0}_{PRPA}$.

Introducing the normalized amplitudes,
\[
{\cal X}_{\mu}(pn;\alpha J)= N_{\mu}^{1/2}(pn) X_{\mu}(pn;\alpha J);\;\;
{\cal Y}_{\mu}(pn;\alpha J)= N_{-\mu}^{1/2}(pn) Y_{\mu}(pn;\alpha J),
\label{36}
\]
the PRPA equation takes the standard form
\[
\left(\begin {array}{cc}
{\cal A}_{\mu}(J) &  {\cal B}(J) \\
-{\cal B}^{\d}(J) &- {\cal A}_{-\mu}^{\ast}(J)
\end {array} \right)
\b{{\cal X}_{\mu}(\alpha J)} {{\cal Y}_{\mu}(\alpha J)}=E_{\mu}(\alpha J)
\b{{\cal X}_{\mu}(\alpha J)} {{\cal Y}_{\mu}(\alpha J)},
\label{37}
\]
with
\begin{eqnarray*}
&&{\cal A}_{\mu}(pn,p'n';J)=\frac{A_{\mu}(pn,p'n';J)}
{\sqrt{N_{\mu}(pn)N_{\mu}(p'n')}},\\
&&{\cal B}(pn,p'n';J)=\frac{B(pn,p'n';J)}{\sqrt{N_{\mu}(pn)N_{-\mu}(p'n')}}.
\label{38}
\end{eqnarray*}

The orthogonality and closure relations are, respectively,
\begin{eqnarray*}
&&{\cal X}_{\mu}^{\d}(\alpha){\cal X}_{\mu}(\beta)
-{\cal Y}_{\mu}^{\d}(\alpha){\cal Y}_{\mu}(\beta)=
 \left\{
\begin{array}{l}
\delta_{\alpha\beta}\;\; \mbox{for}\;E_{\mu}(\alpha)>0\;,\\
-\delta_{\alpha\beta}\;\; \mbox{for}\;E_{\mu}(\alpha)<0\;.\\
\end{array}\right.
\label{39}
\end{eqnarray*}
and:
\begin{eqnarray*}
&&\sum_{E_{\mu}(\alpha)>0}[{\cal X}_{\mu}(\alpha) {\cal X}^{\d}_{\mu}(\alpha)
-{\cal Y}_{-\mu}^{\ast}(\alpha)\tilde{{\cal Y}}_{-\mu}(\alpha)]={\bf 1},
\nonumber\\
&&\sum_{E_{\mu}(\alpha)>0}[{\cal X}_{\mu}^{\ast}(\alpha) \tilde{{\cal
 Y}}_{\mu}(\alpha)
- {\cal Y}_{-\mu}(\alpha){\cal X}_{-\mu}^{\d}(\alpha)]
=0,
\label{40}
\end{eqnarray*}
while the transition amplitudes read
\be
\Lambda_{\mu}(\alpha J)=N_0^{-1/2}\sum_{pn}[
\Lambda_{\mu}^0(pnJ) N_{\mu}^{1/2}(pn) {\cal X}_{\mu}^{\ast}(pn;\alpha J)
+\Lambda_{-\mu}^0(pnJ) N_{-\mu}^{1/2}(pn) {\cal Y}_{\mu}^{\ast}(pn;\alpha J)],
\label{41}
\ee
and for $J^{\pi}=0^+$ and $1^+$
satisfy the sum-rule \cite{lan}
\[
\sum_{E_{\mu=1}(\alpha J)>0}\left|\Lambda_{\mu}(\alpha J)\right|^2
-\sum_{E_{\mu=-1}(\alpha J)>0}\left|\Lambda_{-\mu}(\alpha J)\right|^2
=(2J+1)(N-Z).
\]

The explicit results for the projected RPA matrices are:
\begin{eqnarray*}
&&A_{\mu}(pn,p'n';J)=\epsilon^{Z-1+\mu,N-1-\mu}(pn) \delta_{pn,p'n'}
+R_{22}^{Z-1+\mu,N-1-\mu}(pnp'n';J) ,\nonumber\\
&&B(pn,p'n';J) =R_{40}^{Z,N}(pn,p'n';J),\label{42}
\end{eqnarray*}
where
\begin{eqnarray}
\epsilon^{k,k'}(pn)&=&
[R_0^{k}(p)+R_{11}^{k}(pp)]I^{k'}(n)+[R_0^{k'}(n)+R_{11}^{k'}(nn)]I^{k}(p)
\nonumber\\
&+&R_0^{k,k'}(pn)+R_{11}^{k,k'}(pn) -E_0I^{k}(p)I^{k'}(n),
\label{43}
\end{eqnarray}
are the unperturbed energies, and
\be
E_0=\frac{R_0^{Z}}{I^{Z}}+\frac{R_0^{N}}{I^{N}}+\frac{R_0^{Z,N}}{I^Z I^{N}}.
\label{44}
\ee
is the ground state energy.
The norms are:
\[
N_0=I^ZI^N;\,\,\,N_{\mu}(pn)=I^{Z-1+\mu}(p)I^{N-1-\mu}(n),
\]
and the remaining quantities are defined as follows:
\begin{eqnarray*}
R_0^k(p_1\cdot\cdot)& =&
\sum_p \hat{j}_p^2 v_p^2 e_pI^{k-2}(pp_1\cdot\cdot )\\
&+&\frac{1}{4}\sum_{pp'}\hat{j}_{p}\hat{j}_{p'}
[2v_{p}^2v_{p'}^2{\rm f}(pp') I^{k-4}(pp'p_1\cdot\cdot)
+u_{p}v_{p}u_{p'}v_{p'}{\rm g}(pp')
I^{k-2}(pp'p_1\cdot\cdot)] ,
\end{eqnarray*}
\begin{eqnarray*}
R_{11}^k(p_1p_2\cdot\cdot )& =&
e_{p_1}[u_{p_1}^2 I^{k}(p_1p_2\cdot\cdot)-v_{p_1}^2
I^{k-2}(p_1p_2\cdot\cdot)] \\
&+&\hat{j}_{p_1}^{-1}\sum_{p}\hat{j}_{p}\{v_{p}^2 {\rm f}(pp_1)
[u_{p_1}^2I^{k-2}(pp_1p_2\cdot\cdot )-v_{p_1}^2
I^{k-4}(pp_1\cdot\cdot)]\\
&-&u_pv_pu_{p_1}v_{p_1}{\rm g}(pp_1)I^{k-2}(pp_1p_2\cdot\cdot)\},
\end{eqnarray*}
\[
R_0^{k,k'}(p_1\cdot\cdot ,n_1\cdot\cdot )=
\sum_{pn}\hat{j}_p\hat{j}_n v_p^2v_n^2{\rm f}(pn)
I^{k-2}(pp_1\cdot\cdot)I^{k'-2}(nn_1\cdot\cdot)),
\]
\begin{eqnarray*}
R_{11}^{k,k'}(pn)& =&
\hat{j}_p^{-1}[u_p^2I^{k}(pp)-v_p^2I^{k-2}(pp)]
\sum_{n'}\hat{j}_{n'}v_{n'}^2 {\rm f}(pn')I^{k'-2}(nn')\\
&+&\hat{j}_n^{-1}[u_n^2I^{k'}(nn)-v_n^2I^{k'-2}(nn)]
\sum_{p'}\hat{j}_{p'}v_{p'}^2 {\rm f}(np')I^{k-2}(pp'),
\end{eqnarray*}
\begin{eqnarray*}
R_{22}^{k,k'}(pn,p'n';J)& =&
[u_pv_nu_{p'}v_{n'}I^{k}(pp')I^{k'-2}(nn')
+v_pu_nv_{p'}u_{n'}I^{k-2}(pp')I^{k'}(nn')] {\rm F}(pn,p'n';J)\\
&+& [u_pu_nu_{p'}u_{n'}I^{k}(pp')I^{k'}(nn')+
v_pv_nv_{p'}v_{n'} I^{k-2}(pp')I^{k'-2}(nn')]{\rm G}(pn,p'n';J)\\
R_{40}^{Z,N}(pn,p'n';J)&=&I^{Z-2}(pp') I^{N-2}(nn')
\left[(\vp \un \upp \vnp + \up \vn \vpp \unp ) {\rm F}(pn,p'n';J)\right.\\
&-& \left.(\up \un \vpp \vnp + \vp \vn \upp \unp ) {\rm G}(pn,p'n';J) \right].
\end{eqnarray*}
where $e_p$ are the proton single particle energies (s.p.e.),
${\rm f}(pp')\equiv {\rm F}(pp,p'p';J=0)$,
${\rm g}(pp')\equiv {\rm G}(pp,p'p';J=0)$,
${\rm f}(np)\equiv {\rm F}(nn,pp;J=0)$ and
\[
I^{k}(p_1p_2\cdot\cdot p_n)=
\frac{1}{2\pi i}\oint \frac{dz_{\rm p}}{z_{\rm p}^{k+1}}
\sigma_{p_1}\sigma_{p_2}\cdot\cdot \sigma_{p_n}
\prod_{p}(u_p^2+ z_{\rm p}^2 v_p^2)^{\hat{j}_p^2/2}.
\]

The starting point of a self-consistent numerical work is the
minimization of $E_0$ with regard to the parameters $u$ and $v$. This leads to
a set of coupled equations \cite{die,ott}:
\be
2\hat{e}_p u_pv_p-\Delta_p(u_p^2-v_p^2)=0,
\label{48}
\ee
where the quantities
$\hat{e}_p$ and $\Delta_p$ are defined as follows:
\begin{eqnarray*}
&&\Delta_p=-\frac{1}{2}\hat{j}_p^{-1}\sum_{p'}\hat{j}_{p'}\upp \vpp {\rm
g}(pp')
I^{Z-4}(p), \\
&&\hat{e}_p=\bar{e}_p I^{Z-2}(p)
+\hat{j}_p^{-1}\sum_{p'}\vpp^2 {\rm f}(pp')I^{Z-4}(pp')\\
&+&\frac{1}{2}\hat{j}_p^{-2}\sum_{p'}\hat{j}_{p'}^{2}\vpp^2\bar{e}_{p'}
\left[\nu_p(p')D^{Z-4}(pp')
-\nu_pD^{Z-2}(p')
\frac{I^{Z-2}(p)}{I^Z}\right]\\
&+&\frac{1}{8}\hat{j}_p^{-2}\sum_{p'p''}\hat{j}_{p'}\hat{j}_{p''}
\left\{2v_{p'}^2v_{p''}^2{\rm f}(p'p'')
\left[\nu_p(p'p'')D^{Z-6}(pp'p'')
-\nu_pD^{Z-2}(p)\frac{I^{Z-4}(p'p'')}{I^{Z}}\right] \right.\\
&&+\left.u_{p'}v_{p'}u_{p''}v_{p''}{\rm g}(p'p'')
\left[\nu_p(p'p'')D^{Z-4}(pp'p'')
-\nu_pD^{Z-2}(p)\frac{I^{Z-2}(p'p'')}{I^{Z}}\right]\right \}
\end{eqnarray*}
with
\begin{eqnarray*}
&&\bar{e}_p=e_p+\hat{j}_p^{-1}\sum_n\hat{j}_n \vn^2{\rm f}(pn)
I^{N-2}(n)/I^N,\\
&&\nu_p(p'\cdot\cdot)=\hat{j}_p^2-2(\delta_{pp'}+\cdot\cdot ),\\
&&D^{k}(p\cdot\cdot)= I^{k}(p\cdot\cdot)- I^{k+2}(p\cdot\cdot).
\end{eqnarray*}

To inquire into the effect of the number projection method we have done
numerical calculations for several $2\nu \beta\beta$ emitting nuclei,
following the procedure outlined in ref. \cite{hir}.
This implies the employment of the s.p.e. from the neighboring odd-odd nuclei,
and therefore all self-energy corrections in eqs. (\ref{43}), (\ref{44}) and
(\ref{48}), for both like and unlike particles, have been neglected.
Besides, and for the sake of comparison with results presented in ref.
\cite{tub}, the $0^+\go 0^+$ $2\nu$ transition amplitude has been
approximated as
\br
&&{\cal M}_{2\nu}=\frac{1}{2}\sum_{\alpha,\alpha'}
\left(\frac{1}{E_{\mu=1}(\alpha J=1)+\Delta}
+\frac{1}{\overline{E}_{\mu=-1}(\alpha' J=1)-\Delta}\right)\nonumber\\
&&\times \Lambda_{\mu=1}(\alpha J=1) \ov {\alpha J=1;\mu=1}
{\overline{\alpha' J=1;\mu=-1}} \overline{\Lambda}_{\mu=-1}(\alpha' J=1),
\label{45}
\er
where unbarred and barred quantities indicate that the quasiparticles and
excitations are defined with respect to the initial ($Z$, $N$) nucleus,
and final ($Z+2$, $N-2$) nucleus, respectively, and
$\Delta=\left(E_0- \overline{E}_0 \right)/2$.

As an example, the PQRPA and QRPA calculations for the $^{76}Ge \go {^{76}Se}$
transition are displayed in fig. \ref{fig}. Also, in the same figure
it is shown the result of a hybrid model calculation, in which the PQRPA
is used, but with the BCS parameters $u$ and $v$ employed within the QRPA.
It can be observed that all the three results are qualitative similar, in
the sense that the $\beta\beta$ moments always vary rather abruptly in the
physically relevant interval  $t_0 \gsim t \gsim t_1$,
and are well represented by eq. (\ref{1}); the corresponding parameters
$t_0$ and $t_1$ that fit the exact results, together with
${\cal M}_{2\nu} (t=0)$, are listed in table \ref{tab}.
The single mode model for the $\beta\beta$ decay \cite{krm,krm2} suggests that
the dominant effect of the projection method is to insert, into the numerator
and the denominator of eq. (\ref{1}), terms of the order of
$\left(\Omega^{-1}_p+\Omega^{-1}_n\right)$, where $2\Omega_p$ and $2\Omega_n$
are, respectively, the degeneracy of the proton and neutron levels.

We end the present letter by mentioning  the main differences between the
PRPA formalism reported in ref. \cite{tub} and the one presented here:\\
a) They solve two different RPA equations for the ($Z+1$, $N-1$) and
($Z-1$, $N+1$) nuclei. Contrarily, we only deal with one RPA problem,
in which the $\beta^{+}$ spectrum is viewed as an extension of the $\beta^{-}$
spectrum to negative energies \cite{lan,boh,krm1}.\\
b) Their matrices $A$ and $B$ are defined through the method of commutation
relations. Yet, we feel that this technique cannot be applied with
the projected procedure, as it automatically excludes
the $R_0$ terms in the expression (\ref{43}) for the unperturbed energies.
But these terms in no way cancel among themselves \cite{ott,all}, and thus,
our unperturbed energies are necessarily distinct from those obtained in ref.
\cite{tub}.
(It is worth noting that the commutation method has neither  been used in the
derivation of the PRPA for like particles \cite{fed}.)\\
c) Their result for the matrix $B$ also disagrees with our result and we do
not find any explanation for such a discrepancy. In our formalism the ground
state correlations always occur in the ($Z$, $N$) nucleus.\\
d) We do not have to deal with spurious components in the wave functions of
the intermediate double-odd nucleus as it apparently happens in
ref. \cite{tub}, and therefore $A_{\mu=1},\,A_{\mu=-1}$ and $B$  are square
 matrices of
dimension equal to the number of proton-neutron states with a given spin and
parity.\\
e) The matrix element for the one-body transition operator,
given by eq. (\ref {41}), has not been explicitly shown in their previous
works.

In summary, the QRPA and PQRPA yield qualitatively similar results,
and, in spite of above mentioned differences between the formalism employed
by the T\"{u}bingen group and ours,
the numerical results that we get for the moments ${\cal M}_{2\nu}$ are not
in essence distinct from those obtained in ref. \cite{tub}.

\newpage

\bigskip

\begin{figure}
\begin{center} { \large Figure Captions} \end{center}
\caption{ Calculated double beta decay matrix elements ${\cal M}_{2\nu}$
(in units of $[MeV]^{-1}$) for $^{76}Ge$, as a function of the
particle-particle $S=1$, $T=0$ coupling constant $t$.
Solid and dotted curves correspond to the projected (PQRPA)
and unprojected (QRPA) results, respectively. For comparison a hybrid
model result, in which the PQRPA is used with parameters $u$ and $v$
determined as in the QRPA, is also presented (dashed curve).}
\label{fig}
\end{figure}

\begin{table}
\begin{center} { \large Tables }
\caption{The coefficients ${\cal M}_{2\nu}(t=0)$, $t_0$ and
$t_1$ for $^{76}Ge$ in the parametrization of $2\nu$ moments by eq.
(\protect \ref{1}).
The exact curves, shown in fig. \protect \ref{fig}, are not
distinguished visually from the fitted ones with the parameters
listed here.}
\label{tab}

\bigskip

\begin{tabular}{c|ccc}\hline
&QRPA&PQRPA&PQRPA\\
&&&(hybrid)\\
\hline
$-{\cal M}_{2\nu} (t=0)$    &0.385&0.370&0.297 \\
$t_0$    &1.170&1.038& 0.950 \\
$t_1$    &1.591& 1.447&1.421\\
\hline
\end{tabular}
\end{center}
\end{table}
\end{document}